\documentclass[nobibnotes, prl, aps, letterpaper, nofootinbib, preprint, amsfonts, amssymb, amsmath, superscriptaddress]{revtex4-2}

\usepackage{mathtools}
\usepackage{ragged2e}
\usepackage{color}
\usepackage{bm}
\usepackage[version=4]{mhchem}
\usepackage{hyperref}
\usepackage{xcolor}

\begin{document}

\title{Striped Electronic Phases in an Incommensurately \\ Modulated van der Waals Superlattice}

\author{A. Devarakonda}
\altaffiliation[Present address: ]{Department of Applied Physics and Applied Mathematics,\\ Columbia University, New York, NY, USA}
\affiliation{Department of Physics, Massachusetts Institute of Technology, Cambridge, MA, USA}

\author{A. Chen}
\affiliation{Department of Physics, Massachusetts Institute of Technology, Cambridge, MA, USA}

\author{S. Fang}
\affiliation{Department of Physics, Massachusetts Institute of Technology, Cambridge, MA, USA}

\author{D. Graf}
\affiliation{National High Magnetic Field Laboratory, Tallahassee, FL, USA}

\author{M. Kriener}
\affiliation{RIKEN Center for Emergent Matter Science (CEMS), Wako, Japan}

\author{A. J. Akey}
\affiliation{Center for Nanoscale Systems, Harvard University, Cambridge, MA, USA}

\author{D. C. Bell}
\affiliation{Center for Nanoscale Systems, Harvard University, Cambridge, MA, USA}
\affiliation{John A. Paulson School of Engineering and Applied Sciences, Harvard University, Cambridge, MA, USA}

\author{T. Suzuki}
\affiliation{Department of Physics, Toho University, Funabashi, Japan}

\author{J. G. Checkelsky}
\affiliation{Department of Physics, Massachusetts Institute of Technology, Cambridge, MA, USA}

\maketitle

\clearpage

\setlength{\parindent}{4ex}
\noindent
\justify

\textbf{The electronic properties of crystalline materials can be manipulated by superimposing spatially periodic electric, magnetic, or structural modulations. Systems with long-wavelength modulations incommensurate with the atomic lattice are of particular interest \cite{cumminsExperimentalStudiesStructurally1990}, exemplified by recent advances with moir\'e patterned two-dimensional (2D) van der Waals (vdW) heterostructures \cite{bistritzerMoireBandsTwisted2011,caoCorrelatedInsulatorBehaviour2018}. Bulk vdW superlattices \cite{huVanWaalsAntiferromagnetic2020,devarakondaClean2DSuperconductivity2020,devarakondaSignaturesBosonicLandau2021,maTwodimensionalSuperconductivityBulk2022,perskyMagneticMemorySpontaneous2022} hosting 2D interfaces between minimally disordered layers represent scalable bulk analogs of artificial vdW heterostructures and present a complementary venue to explore incommensurately modulated 2D states.  Here we report the bulk vdW superlattice \ce{SrTa2S5} realizing an incommensurate one-dimensional (1D) structural modulation of 2D transition metal dichalcogenide (TMD) $H$-\ce{TaS2} layers. High-quality electronic transport in the $H$-\ce{TaS2} layers, evidenced by quantum oscillations, is made anisotropic by the modulation and exhibits commensurability oscillations \cite{beenakkerGuidingcenterdriftResonancePeriodically1989} paralleling lithographically modulated 2D systems \cite{gerhardtsNovelMagnetoresistanceOscillations1989,ensslinMagnetotransportAntidotLattice1990,huberBandConductivityOscillations2022}. We also find unconventional, clean-limit superconductivity (SC) in \ce{SrTa2S5} with a pronounced suppression of interlayer coherence relative to intralayer coherence. Such a hierarchy of energy scales can arise from pair-density wave (PDW) SC with mismatched spatial arrangement in adjacent superconducting layers \cite{liTwodimensionalSuperconductingFluctuations2007, bergDynamicalLayerDecoupling2007,tranquadaCuprateSuperconductorsViewed2020}. Examining the in-plane magnetic field ($H_{ab}$) dependence of interlayer critical current density $J_c$, we observe anisotropy with respect to $H_{ab}$ orientation: $J_c$ is maximized (minimized) when $H_{ab}$ is perpendicular (parallel) to the stripes, consistent with 1D PDW SC \cite{yangDetectionStripedSuperconductors2013, yangJosephsonEffectFuldeFerrellLarkinOvchinnikov2000,lozanoTestingPairDensity2022}. Electron diffraction shows the structural stripe modulation is phase-shifted between adjacent $H$-\ce{TaS2} layers, suggesting mismatched 1D PDW is seeded by the striped structure. Hosting a high-mobility Fermi liquid in a coherently modulated structure, \ce{SrTa2S5} is a promising venue for unusual phenomena anticipated in clean, striped metals and superconductors \cite{neilsExperimentalTestSubdominant2002,agterbergDislocationsVorticesPairdensitywave2008,bergCharge4eSuperconductivityPairdensitywave2009}. More broadly, \ce{SrTa2S5} establishes bulk vdW superlattices as versatile platforms to address long-standing predictions surrounding modulated electronic phases in the form of macroscopic crystals \cite{tranquadaEvidenceStripeCorrelations1995,vallaGroundStatePseudogap2006}}.

Incommensuration in crystals arises from the coexistence of two or more mutually incompatible periodicities \cite{axeIncommensurateStructures1980}. Incommensuration can appear in forms spanning magnetic modulations in rare earth metals \cite{overhauserExchangeCorrelationInstabilities1968,izyumovModulatedLongperiodicMagnetic1984}, arrangement of ions and molecules in graphite intercalation compounds \cite{dresselhausIntercalationLayeredMaterials1986}, and incommensurate charge density waves (CDWs) in metallic systems \cite{monceauElectronicCrystalsExperimental2012}. Another prominent class is structurally incommensurate materials where incommensuration appears as a spatial modulation of the crystal lattice. These materials, with mismatched periodicities of the lattice and incommensurate modulation, are beyond the usual description of crystals and motivated the formulation of superspace group methods to describe their structures \cite{dewolffSuperspaceGroupsIncommensurate1981}. If the incommensurate modulation is weak and has a longer wavelength compared to the lattice periodicity, it can act as a perturbation to the underlying crystal \cite{esakiSuperlatticeNegativeDifferential1970}, presenting a means to manipulate their electronic behavior and design phases beyond those found in conventional crystals.

\begin{figure*}[h!]
    \makebox[\textwidth][c]{\includegraphics[scale=1.0]{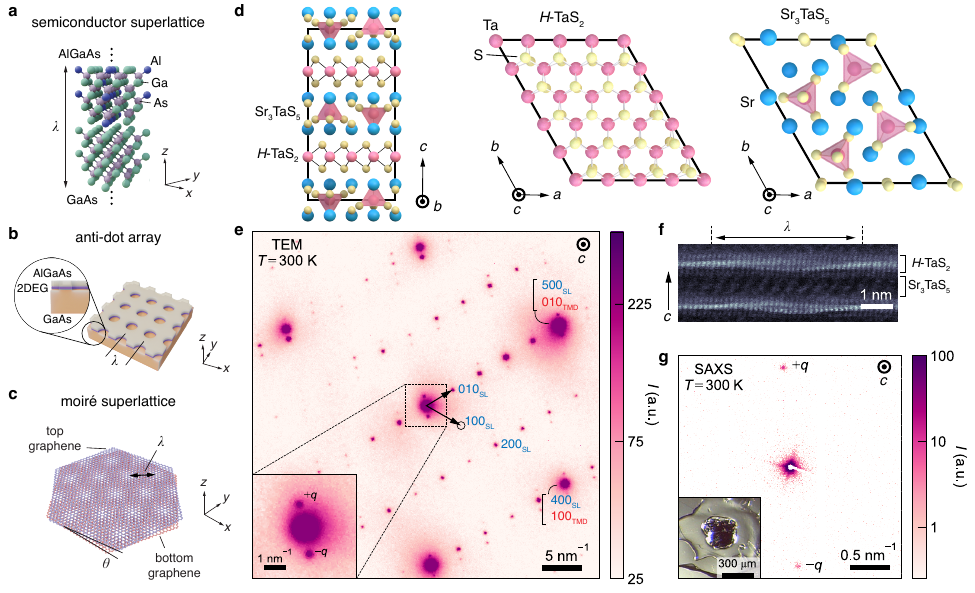}}
\caption{Devarakonda \textit{et al.}}
\label{fig:1}
\end{figure*}

Semiconductor superlattices \cite{changGrowthGaAsGaAlAs1973} are early examples of engineered incommensuration where one-dimensional (1D) periodic modulations along the growth direction (Fig. \ref{fig:1}a) are created by varying the chemical composition, for example, periodic aluminum (Al) doping into gallium arsenide (GaAs). Incommensuration is achieved therein by engineering the modulation wavelength $\lambda$ such that $\lambda / a$ is irrational ($a$ is the crystal lattice spacing). Lithographically patterned semiconductor heterostructures are another example (Fig. \ref{fig:1}b) with incommensurate modulations superimposed on high-quality 2D electron gases (2DEGs). More recently, moir\'e patterns at lattice mismatched or rotationally faulted vdW heterointerfaces (Fig. \ref{fig:1}c) have emerged as hosts for incommensurately modulated phases. Together, materials with engineered incommensuration have realized various intriguing electronic phenomena: semiconductor superlattices \cite{esakiSuperlatticeNegativeDifferential1970, changGrowthGaAsGaAlAs1973, esakiNewTransportPhenomenon1974} feature electronic mini-bands with non-linear electronic and optical properties, patterned 2DEGs exhibit novel commensurabilty resonances \cite{gerhardtsNovelMagnetoresistanceOscillations1989,ensslinMagnetotransportAntidotLattice1990}, and moir\'e patterned vdW heterostructures host an array of emergent correlated and topological phases \cite{caoCorrelatedInsulatorBehaviour2018,serlinIntrinsicQuantizedAnomalous2020,liQuantumAnomalousHall2021}. A notable aspect of these existing platforms is their exceptionally low atomic-scale disorder, supporting high-mobility transport and fragile electronic ground states. Thermodynamically stable, incommensurately modulated bulk materials with comparable quality are desirable as readily accessible, complementary platforms to study modulated electronic states. Apart from fragile organic conductors \cite{kawamotoOrganicSuperconductorsIncommensurate2009}, modulated bulk materials exhibit relatively low electronic mobilities (Sec. SIII).

Here we report a bulk vdW superlattice \cite{huVanWaalsAntiferromagnetic2020,devarakondaClean2DSuperconductivity2020,maTwodimensionalSuperconductivityBulk2022,perskyMagneticMemorySpontaneous2022} which naturally forms a macroscopically uniform incommensurate structural modulation with pronounced effects on its electronic behavior. The material \ce{SrTa2S5} is composed of $H$-\ce{TaS2} TMD and \ce{Sr3TaS5} spacer layers stacked in an alternating fashion (see Fig. \ref{fig:1}d, left). In the average structure, the latter has a 2-fold symmetric in-plane structure (monoclinic point group $2$) which forms a commensurate $4 \times 5$ superstructure with 3-fold symmetric $H$-\ce{TaS2} (hexagonal point group $\bar{6}m2$) (Fig. \ref{fig:1}d, right and Sec. SIa). Figure \ref{fig:1}e shows an electron diffraction pattern of the $ab$-plane structure where these features can be identified.  We find dominant reflections from $H$-\ce{TaS2} (Fig. \ref{fig:1}e, red) and superstructure reflections from monoclinic \ce{Sr3TaS5} (Fig. \ref{fig:1}e, blue). We also observe satellite reflections at $\pm \vec{q}$ (Fig. \ref{fig:1}e, inset) from a long-wavelength 1D modulation. Using $|\vec{q}| = 2 \pi / \lambda$ we estimate $\lambda \approx 4.4$ nm. The $\vec{q}$-vector has an irrational relationship with the reciprocal lattice of the average structure, evidencing incommensuration (section SIb). High-resolution synchrotron powder x-ray diffraction (PXRD) shows the TMD layers are weakly strained towards commensuration with the spacers to form the $4 \times 5$ superstructure (section SIb). This parallels misfit superlattices \cite{wiegersMisfitLayerCompounds1996,kuypersIncommensurateMisfitLayer1989} and in-plane TMD heterointerfaces \cite{xieCoherentAtomicallyThin2018} where translation symmetry mismatch gives rise to incommensurate structural modulations.

The 1D stripes are apparent in cross-section (Fig. \ref{fig:1}f) and $ab$-plane (see Fig. S3a) transmission electron microscopy (TEM) images. We observe an out-of-plane distortion of the $H$-\ce{TaS2} layers with peak-to-valley amplitude $\delta z \approx 2.3$ \AA\, and lateral periodicity of several nanometers, consistent with $ab$-plane electron diffraction. Cross-section electron diffraction indicates this structural modulation is $\pi$ phase-shifted between adjacent layers (Sec. SIb). Small-angle x-ray scattering (SAXS) at temperature $T = 300$ K from the $ab$-plane of bulk crystals (Fig. \ref{fig:1}g, sample in inset) shows the same $\pm \vec{q}$ pattern of satellite reflections with $\lambda = 4.385 \pm 0.015$ nm (Methods). Given the x-ray beam spot encompasses an area comparable to the measured crystals, the single pair of sharp $\pm\vec{q}$ reflections demonstrate the incommensurate structural modulation is macroscopically coherent. We hypothesize this large scale coherence is rooted in the low symmetry of the average structure and the 1D nature of the modulation which restricts the allowed modulation directions (section SIb).

\begin{figure*}[h!]
    \makebox[\textwidth][c]{\includegraphics[scale=1.0]{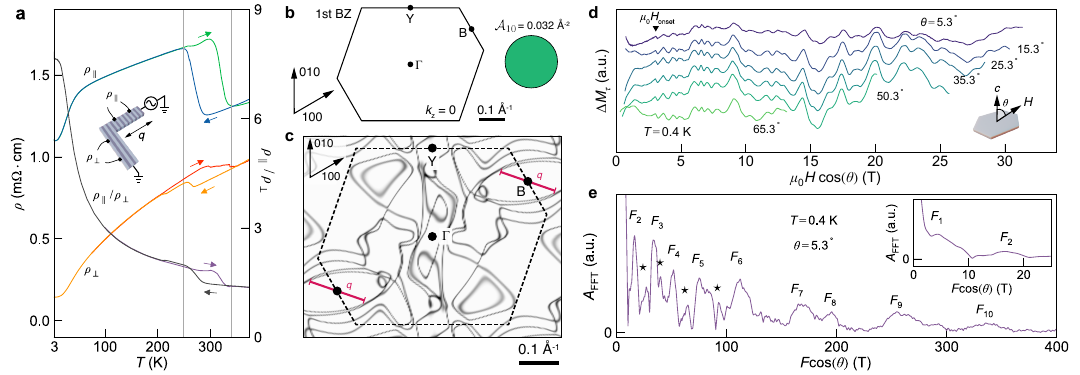}}
\caption{Devarakonda \textit{et al.}}
\label{fig:2}
\end{figure*}

Turning to the electronic properties of \ce{SrTa2S5}, we examined the $T$ dependence of in-plane resistivity perpendicular ($\rho_\perp$) and parallel ($\rho_\parallel$) to $\vec{q}$ using a focused ion beam (FIB) patterned ``L-bar'' device (Fig. \ref{fig:2}a, inset). Similar to the parent TMD 2$H$-\ce{TaS2}, this system is a metal. Both $\rho_\perp \left( T \right)$ (Fig. \ref{fig:2}a, orange and red) and $\rho_\parallel \left( T \right)$ (Fig. \ref{fig:2}a, blue and green) show prominent thermal hysteresis in the range $250 \textnormal{ K} \lesssim T \lesssim 350 \textnormal{ K}$ signaling a first-order phase transition. While synchrotron PXRD, SAXS, and electron diffraction do not evidence a structural transition, we find similar thermal hysteresis in torque magnetometry which points to an electronic origin (section SII). Additionally, first-principles calculations (Fig. \ref{fig:2}c) reveal Fermi surface (FS) segments centered at $\textrm{B}$ which are well-nested by the modulation $q$-vector (Fig. \ref{fig:2}c, red). Together, these point to a first-order lock-in CDW transition as the source of the thermal hysteresis (section SII). At lower $T$ the in-plane transport anisotropy $\rho_\parallel / \rho_\perp$ (Fig. \ref{fig:2}a, gray and purple) increases and reaches $ \rho_\parallel / \rho_\perp \approx 8$ at $T = 3$ K. Comparable $\rho_\parallel / \rho_\perp$ is found in the high-field nematic phase of 2DEGs \cite{fradkinNematicPhaseTwoDimensional2000}, some stripe-phase cuprates \cite{andoElectricalResistivityAnisotropy2002}, and misfit superlattices \cite{sakabayashiCrossoverItinerantLocalized2021}, suggesting the stripe modulation acts like a perturbation; the in-plane anisotropy is markedly less than that of TMDs such as $1T'$-\ce{WTe2} with quasi-1D intralayer structures \cite{hoInplaneAnisotropyOptical2001,jhaAnisotropyElectronicTransport2018}.

We also observe quantum oscillations in \ce{SrTa2S5} that allow us to map its fermiology. Figure \ref{fig:2}d shows de Haas-van Alphen (dHvA) oscillations in torque magnetization $\Delta M_\tau (1/H_\perp)$ (section SIII) at $T = 0.4$ K versus the magnetic field component perpendicular to the $ab$-plane, $H_\perp = H \cos \theta$; $H$ is the total field and $\theta$ is measured from the $c$-axis (Fig. 2d, inset). Alignment of $\Delta M_\tau (1/H_\perp)$ across various fixed $\theta$ indicates the FSs are cylindrical and aligned along $c$ due to weak interlayer coupling. The complex oscillatory behavior of $\Delta M_\tau \left( H \right)$ is due to interference amongst numerous oscillation frequencies and corresponding FSs. We extract the individual frequency $F$ components from the Fast Fourier Transform (FFT) of $\Delta M_\tau \left( 1 / H_\perp \right)$. The FFT spectrum for $\theta = 5.3^\circ$ and $T = 0.4$ K is shown in Fig. \ref{fig:2}e, exhibiting numerous $F$ components spanning $F_1 = 4.4$ T to $F_{10} = 335$ T (see Sec. SIII), reflecting the complex electronic structure of \ce{SrTa2S5} (Fig. \ref{fig:2}c). We also find weaker $F$ contributions (Fig. \ref{fig:2}e, starred) which we ascribe to magnetic breakdown between closely spaced Fermi surfaces (Sec. SIII \cite{falicovMagnetoresistanceMagneticBreakdown1964}), further consistent with the complex electronic structure (Fig. \ref{fig:2}c).

Using Onsager's relation (Methods) \cite{shoenbergMagneticOscillationsMetals1984} we find the largest FS of \ce{SrTa2S5} encloses an area $\mathcal{A}_{10} = 0.032$ \AA$^{-2}$ (Fig. \ref{fig:2}b) in the Brillouin zone (BZ), $10\times$ smaller than anticipated for $H$-\ce{TaS2} (section SIII). This is consistent with zone-folding of the $H$-\ce{TaS2} FSs into a nearly $20\times$ smaller Brillouin zone (BZ) of \ce{SrTa2S5} due to the $4 \times 5$ superlattice periodicity imposed by the spacer. More broadly, the appearance of quantum oscillations in this incommensurately modulated material is remarkable. From their onset field $\mu_0 H_\textnormal{onset} \approx 3$ T (Fig. \ref{fig:2}d, black triangle) we estimate a quantum mobility $\mu_\textnormal{q} \approx$ 3,000 cm$^2$/Vs, markedly larger than mobilities seen in 2$H$-\ce{TaS2} ($\mu \approx 1$ cm$^2$/Vs \cite{naitoElectricalTransportProperties1982}) and $H$-\ce{TaS2} containing misfit-layer compounds ($\mu \approx 25$ cm$^2$/Vs \cite{suzukiOpticalReflectivityCarrier1995}, Sec SIII). Beyond presenting a bulk platform for exploring phase-coherent electrons in incommensurate potentials \cite{spurrierTheoryQuantumOscillations2019}, the high quality of \ce{SrTa2S5} may support exotic but fragile modulated states. Notably, this mobility enhancement parallels that found in the related $H$-\ce{NbS2} containing superlattice \ce{Ba6Nb11S28} relative to 2$H$-\ce{NbS2} \cite{devarakondaClean2DSuperconductivity2020}, which suggests this growing family of superlattices generically hosts clean TMD layers.

\begin{figure*}[h!]
    \makebox[\textwidth][c]{\includegraphics[scale=1.25]{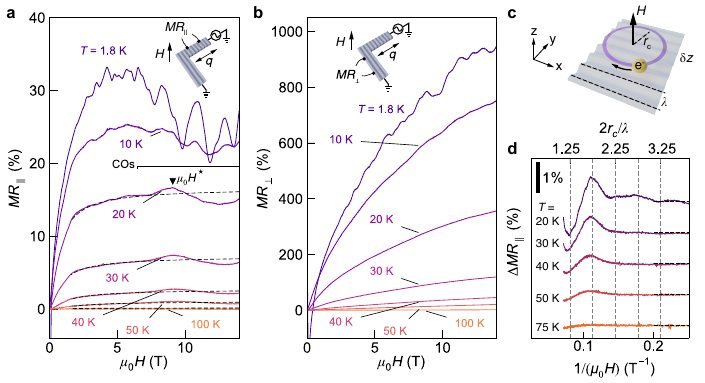}}
\caption{Devarakonda \textit{et al.}}
\label{fig:3}
\end{figure*}

The stripe modulation in \ce{SrTa2S5} leads to additional transport features characteristic of modulated 2DEGs. Figures \ref{fig:3}a and \ref{fig:3}b show magnetoresistance of the L-bar device measured parallel and perpendicular to $\vec{q}$, $MR_\parallel \left( H \right)$ and $MR_\perp \left( H \right)$, respectively (Methods). Together with Shubnikov-de Haas (SdH) oscillations paralleling dHvA oscillations in $\Delta M_\tau$, we find a low frequency $F_\textnormal{CO} \approx 15.5$ T oscillation in $MR_\parallel$ that is absent in $MR_\perp$ (section SIV). Further, this oscillation persists to $T > 50$ K unlike dHvA oscillations which are suppressed by $T \approx 25$ K (Fig. S5). The absence of these oscillations in $MR_\perp$ indicates $\vec{q}$ of the stripes is relevant to their observation. The overall phenomenology parallels commensurability oscillations (COs) in stripe-patterned 2DEGs due to matching of the cyclotron diameter $2r_\textnormal{c} = 2 \hbar k_F / (e \mu_0 H)$ and modulation $\lambda$ (see Fig. \ref{fig:3}c) with frequency $F_\textnormal{CO} = 2 \hbar k_\textnormal{F} / e \lambda$ where $k_\textnormal{F}$ is the Fermi wavevector \cite{beenakkerGuidingcenterdriftResonancePeriodically1989, gerhardtsNovelMagnetoresistanceOscillations1989}. Using the observed $F_\textnormal{CO}$ and $\lambda = 4.38$ nm from SAXS we find $k_\textnormal{F} \approx 6 \times 10^{-3}$ \AA$^{-1}$, in nominal agreement with $k_\textnormal{F} \approx 1 \times 10^{-2}$ \AA$^{-1}$ of the smallest pocket observed in dHvA oscillations (see Sec. SIII). Moreover, these oscillations exhibit the characteristic $1/4$ period phase shift anticipated for commensurability oscillations (Fig. \ref{fig:3}d) \cite{winklerLandauBandConductivity1989, raichevEffectLandauQuantization2018} and distinct from conventional quantum oscillations \cite{shoenbergMagneticOscillationsMetals1984}.

\begin{figure*}[hbt!]
    \makebox[\textwidth][c]{\includegraphics[scale=1.0]{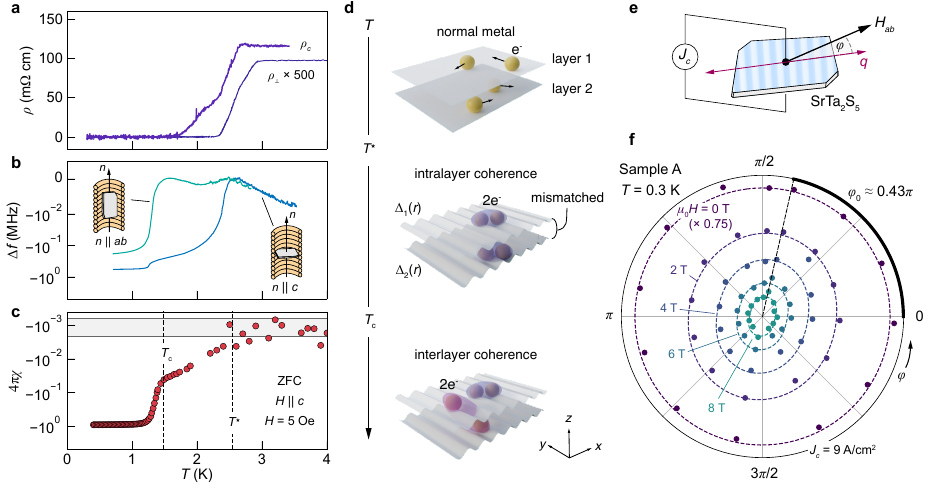}}
    \caption{Devarakonda \textit{et al.}}
    \label{fig:4}
\end{figure*}

At low $T$ superconductivity (SC) emerges from the stripe-modulated metallic state. Comparing the $T$ dependence of intralayer (perpendicular to $\vec{q}$) and interlayer resistivity, $\rho_\perp (T)$ and $\rho_\textnormal{c} (T)$ respectively, we find $\rho_\textnormal{c}$ reaches zero at lower $T$ relative to $\rho_\perp$ (Fig. \ref{fig:4}a), suggesting suppressed interlayer coupling in the superconducting state. We also performed tunnel diode oscillator (TDO) measurements in this low $T$ regime, wherein the $LC$ oscillator frequency $f$ reflects changes in screening supercurrents \cite{schawlowEffectEnergyGap1959,giannettaLondonPenetrationDepth2022} (Methods).  The variation of $f$ below $T$ = 2.5 K, $\Delta f(T) = f(T) - f(2.5 \textnormal{ K})$ for crystals oriented with their $c$-axis or $ab$-plane parallel to the coil axis $n$, sensing the intralayer or interlayer supercurrent response, respectively, exhibit a similar hierarchy of $T$ scales (Fig. \ref{fig:4}b). For the same sample, we observe that intralayer screening onsets near 2.3 K (Fig. \ref{fig:4}b, blue) whereas interlayer screening onsets near 1.4 K (Fig. \ref{fig:4}b, green). Similar separation of $T$ scales is observed in zero-field cooled (ZFC) magnetic susceptibility $\chi (T)$ for $H \parallel c$. Below $T^\star = 2.5$ K where $\chi (T)$ crosses $ \bar{\chi} (3 \textnormal{ K} < T < 5 \textnormal{ K}) \pm \sigma_\chi (3 \textnormal{ K} < T < 5 \textnormal{ K})$ ($\bar{\chi}$ and $\sigma_\chi$ denote the mean and standard deviation, respectively), we observe weak diamagnetism consistent with the appearance of intralayer supercurrents (Fig. \ref{fig:4}c). This is followed by a Meissner transition at $T_c = 1.49$ K with a shielding fraction $4 \pi \chi$ near unity, characteristic of bulk superconductivity (Sec. SVa).

Together these observations establish that intralayer superconducting coherence appears at $T^\star$ followed by interlayer coherence at $T_c$ (\textit{i.e.} superconducting coupling between the layers is frustrated). This sequence of transitions parallels the phenomenology of striped cuprate superconductors such as La$_{1.875}$Ba$_{0.125}$CuO$_4$ \cite{liTwodimensionalSuperconductingFluctuations2007,tranquadaCuprateSuperconductorsViewed2020}. In these cuprates, interlayer coherence is thought to be suppressed by pair-density wave (PDW) SC with spatially modulated superconducting order parameter $\Delta(\vec{r})$ that is mismatched between adjacent \ce{CuO2} layers \cite{bergDynamicalLayerDecoupling2007, himedaStripeStatesSpatially2002}. Recent scanning tunneling microscopy (STM) experiments \cite{liuDiscoveryCooperpairDensity2021,guDetectionPairDensity2023,liuPairDensityWave2023} indicate PDW order is a generic feature of SC emerging from modulated metals, raising \ce{SrTa2S5} as a host for 1D PDW SC with $\Delta \sim \sin(\vec{q}\cdot\vec{r})$. A natural hypothesis is that the inequivalence of the structural modulation in adjacent layers (Sec. SIb) seeds similarly inequivalent $\Delta(\vec{r})$. Therefrom, numerous configurations of 1D PDW order mismatched between neighboring layers can suppress interlayer coherence (Sec. SVb) and potentially lead to the observed separation of energy scales (Fig. \ref{fig:4}d).

To test this hypothesis, we study the interlayer critical current density $J_c$ versus in-plane magnetic field strength $H_{ab}$ and orientation $\varphi$ relative to $\vec{q}$ (Fig. \ref{fig:4}e). For finite $H_{ab}$ oriented along $\varphi \approx \pi/2$, suppression of $J_c$ by mismatched PDW order is partially compensated by the Lorentz boost of Cooper pairs from $H_{ab}$, which is a global phase-sensitive consequence of PDW SC \cite{yangDetectionStripedSuperconductors2013, yangJosephsonEffectFuldeFerrellLarkinOvchinnikov2000} (Sec. SVc). Correspondingly, $J_{c} (\varphi)$ takes a characteristic, anisotropic form with $J_{c}(\varphi \approx \pi/2)$ enhanced relative to $J_{c}(\varphi \approx 0)$ (Sec. SVc) \cite{yangJosephsonEffectFuldeFerrellLarkinOvchinnikov2000,yangDetectionStripedSuperconductors2013,lozanoTestingPairDensity2022}. Figure \ref{fig:4}f shows $J_{c}(\varphi)$ at $T = 0.3$ K for various fixed $|H_{ab}|$; with increasing field, a pronounced 2-fold anisotropy develops. Quantitatively, least-squares fits (Fig. \ref{fig:4}f, dashed lines) to $J_c(H_{ab}, \varphi) = J_c^0(H_{ab}) + J_c^1(H_{ab})\cos^2(\varphi + \varphi_0)$ (the second term captures partial restoration of $J_c$ by $H_{ab} (\varphi)$) reveals $ \varphi_0 = 0.43\, \pi $. Similar behavior is also observed in a second sample (Sec. SVc). The observed $\varphi_0$ is consistent with $\varphi_0 = \pi/2$ anticipated for 1D PDW SC seeded by the structural stripe modulation.

\ce{SrTa2S5} is in the clean-limit of superconductivity (Sec. SVa), raising it as a host for various unconventional phases anticipated in clean PDW superconductors \cite{agterbergDislocationsVorticesPairdensitywave2008,smidmanSuperconductivitySpinOrbit2017}. Further, the homogeneity of the incommensurate modulation suggests the PDW is macroscopically uniform which is advantageous for further examination. Also, unlike the cuprates, PDW SC in \ce{SrTa2S5} emerges from a high-mobility, stripe-modulated Fermi liquid, making \ce{SrTa2S5} a potential model system to study this phase. Signatures of PDW SC in \ce{SrTa2S5} also lends further support to the notion that PDWs are a generic feature of SC in modulated metals \cite{liuDiscoveryCooperpairDensity2021,guDetectionPairDensity2023,liuPairDensityWave2023}. Beyond addressing long-standing questions surrounding unconventional, modulated superconductivity, this growing family of bulk vdW superlattices \cite{devarakondaClean2DSuperconductivity2020,maTwodimensionalSuperconductivityBulk2022} may chart a path towards scaling moir\'e heterostructures and their analogs up to macroscopic scales.

\section{Methods}

\noindent \textbf{Single crystal synthesis} Single crystals of \ce{SrTa2S5} were obtained from a high-temperature reaction of SrS, Ta, and S in the presence of \ce{SrCl2} (all powders with purities $\geq 99.99\% $). Thin, lustrous black plates with typical dimension $0.4 \times  0.4 \times 0.03$ mm$^3$ (length $\times$ width $\times$ thickness) were extracted from the product. The crystal structure was analyzed by powder x-ray diffraction (PXRD), electron diffraction, and high-angle annular dark-field scanning transmission electron microscopy (HAADF-STEM). In-house PXRD was performed using a commercial diffractometer with a Cu source. High-resolution synchrotron PXRD data was collected at beamline 11-BM at the Advanced Photon Source (APS), Argonne National Laboratory at an average $\lambda = 0.458104$ \AA. Samples for in-house and synchrotron PXRD were prepared by grinding crystals together with amorphous silica to mitigate preferred orientation effects and reduce absorption by the heavy constituent elements.

\medskip

\noindent \textbf{Small-angle x-ray scattering (SAXS)} SAXS experiments in transmission geometry were performed using a commercial SAXSLAB system equipped with a variable temperature sample holder ($T = 400$ K to $100$ K). Samples were affixed to thin mica membranes using GE varnish. We used an x-ray beam spot approximately 250 $\mu$m in diameter for all measurements. The uncertainty in computing $\lambda$ is due to misalignment of the sample $c$-axis with respect to the incoming beam that makes $|+q|\neq|-q|$.

\medskip

\noindent \textbf{Scanning transmission electron microscopy (STEM)} STEM experiments were conducted at a CEOS Cs probe corrected cold emission gun JEOL JEM-ARM200F STEM operated at 200 kV acceleration voltage. HAADF-STEM images were acquired with 75 mrad convergence semi-angle and 2D Wiener filter applied to reduce the noise. Samples were prepared by a FEI Helios focused ion beam, operated at 30 kV acceleration voltage for the gallium beam during lift-out and 2 kV during polishing. Additional polishing was performed at 0.5 kV with a Fischione NanoMill for 10 minutes on each side at a milling angle of $\pm 10^\circ$.

\medskip

\noindent \textbf{Focused ion beam (FIB) sample preparation} ``L-bar" devices for transport anisotropy measurements were prepared using an FEI Helios focused ion beam system (gallium ion source) operating at 30 kV. Coarse milling was performed using a 10 nA beam current followed by side-wall polishing using a 1 nA beam current.

\medskip

\noindent \textbf{Transport measurements}
Longitudinal and transverse resistivity were measured using standard AC lock-in techniques. Longitudinal (transverse) voltages were (anti-)symmetrized to correct for contact misalignment. We compute $MR_\perp(H)$ and $MR_\parallel(H)$ in the superconducting state for $T \lesssim 3$ K using $\rho_\perp \left( 0 \right)$ and $\rho_\parallel \left( 0 \right)$ in the normal state at $T = 3$ K. Current-voltage $I(V)$ characteristics were measured in a 4-probe configuration. The sample was voltage biased (Yokogawa GS200) and the resulting current was measured using a current pre-amplifier (DL Instruments Model 1211). The longitudinal voltage was simultaneously measured using a Keithley 2182A nanovoltmeter. Up- and down-sweep $I(V)$ traces were coincident, indicating Joule heating is minimal. The in-plane magnetic field dependence of interlayer $I(V)$ was examined at the National High Magnetic Field Lab using a $^3$He cryostat equipped with a $2$-axis rotator probe and superconducting solenoid magnet.

\medskip

\noindent \textbf{Torque and SQUID magnetometry}
Torque magnetometry measurements were performed using commercial, piezoresistive silicon cantilevers (Seiko PRC-400 and SCL Sensortech PRSA-L300). The piezoresistive elements, one with sample mounted and one empty for reference, were incorporated into a Wheatstone bridge and balanced at zero-field. The torque signal was detected using standard AC lock-in techniques with excitations below 20 mV across the bridge. A co-mounted Hall sensor is used to calibrate the rotation angle. Magnetization down to 0.39 K was measured with a SQUID magnetometer in a commercial magnetic property measurement system (Quantum Design MPMS3) equipped with a $^3$He refrigerator. We perform a demagnetization correction to account for the plate-like habit of the samples (see Sec. SVa).

\medskip

\noindent \textbf{Tunnel diode oscillator measurements}
Tunnel diode oscillator (TDO) measurements were performed with an \ce{SrTa2S5} crystal placed within the inductive coil of an $LC$ tank circuit, driven at its resonant frequency $f$ by a tunnel diode based bias circuit (see \cite{vandegriftTunnelDiodeOscillator1975,coffeyMeasuringRadioFrequency2000}). The inductive coil was constructed from 30 turns of 50 AWG (0.025 mm diameter) copper wire. The TDO circuit was biased at a current $I$ = 10 mA and operated at a resonant frequency $f \approx 120$ MHz. A heterodyne circuit was used to down-convert $f$ to the $\mathcal{O}$(1 MHz) range and detect $\Delta f$.

\medskip

\noindent \textbf{Density functional theory calculations} We performed electronic structure calculations implemented in the Vienna ab initio simulation package \cite{Kresse1996,Kresse1996a} using the projector augmented wave pseudo-potential method \cite{Blochl1994} and exchange–correlation functional within the generalized gradient approximation parametrized by Perdew–Burke–Ernzerhof \cite{Perdew1996}.

\medskip

\noindent \textbf{Acknowledgments} We are grateful to E. Kaxiras, S. Y. F. Zhao, and J. P. Wakefield for fruitful discussions. This work was funded, in part, by the Gordon and Betty Moore Foundation FPiOS Initiative, Grant No. GBMF9070 to J.G.C (instrumentation development, DFT calculations), the US Department of Energy (DOE) Office of Science, Basic Energy Sciences, under award DE-SC0022028 (material development), and the Office of Naval Research (ONR) under award N00014-21-1-2591 (advanced characterization).  A.D. acknowledges support from the Simons Foundation, Society of Fellows program (grant No. 855186). D.C.B. acknowledges support from the STC Center for Integrated Quantum Materials (NSF grant DMR-1231319). A portion of this work was performed at the National High Magnetic Field Laboratory, which is supported by the National Science Foundation Cooperative Agreement no. DMR-1644779, the State of Florida and the DOE. Use of the Advanced Photon Source at Argonne National Laboratory was supported by the DOE under Contract No. DE-AC02-06CH11357.

\medskip

\noindent \textbf{Author contributions}  A.D. synthesized and characterized single crystals and fabricated FIB devices. A.D. and D.G. performed the electrical transport and torque magnetometry experiments. A.D., A.C., and D.G. performed tunnel diode oscillator characterization. M.K. performed the SQUID magnetization experiments. A.J.K. and D.C.B. performed the electron microscopy experiments. A.D. performed analytical calculations and S.F. performed electronic structure calculations. A.D. and J.G.C wrote the manuscript with contributions and discussions from all authors. J.G.C. supervised the project.

\bibliographystyle{./nature.bst}

\pagebreak

\noindent\textbf{Periodically modulated crystals and a structurally modulated bulk superlattice}
\\\textbf{a} AlGaAs semiconductor superlattice with Al composition modulated with wavelength $\lambda$ along the $c$-axis. \textbf{b} 2DEG at AlGaAs-GaAs interface patterned with anti-dot array. \textbf{c} Graphene layers rotationally misaligned by angle $\theta$ form a moir\'e pattern. \textbf{d} (left) Inversion symmetric stacking of \ce{Sr3TaS5} and $H$-\ce{TaS2} in \ce{SrTa2S5} (average structure). In-plane structures of (middle) $H$-\ce{TaS2} TMD and (right) \ce{Sr3TaS5} spacer layers. \textbf{e} Electron diffraction pattern of the $ab$-plane at $T = 300$ K shows reflections from $H$-\ce{TaS2} (red), \ce{Sr3TaS5} spacer layers (blue), and satellite reflections from a 1D structural modulation (inset). \textbf{f} Real-space TEM cross-section showing out-of-plane structural modulation of $H$-\ce{TaS2} layers. \textbf{g} SAXS diffraction pattern in the $ab$-plane at $T = 300$ K shows satellite reflections from the structural modulation.

\pagebreak

\noindent\textbf{Transport anisotropy and fermiology of SrTa$_2$S$_5$}
\\\textbf{a} $\rho_\perp\left(T\right)$ (warming (red) and cooling (orange)), $\rho_\parallel\left(T\right)$ (warming (green) and cooling (blue)), and $\rho_\parallel / \rho_\perp \left( T \right)$ (warming (purple) and cooling (gray)) measured in a ``L-bar'' device (inset schematic). \textbf{b} Brillouin zone (BZ) of \ce{SrTa2S5} with the Fermi surface cross-section area of $\mathcal{A}_{10}$ drawn to scale as a circle. \textbf{c} First-principles calculation of the Fermi surfaces in the first BZ (dashed outline). The pocket centered at \textrm{B} is well-nested by the $q$-vector (red) of the structural modulation. \textbf{d} $\Delta M_\tau$ ($H_\perp$) with $H$ inclined at various angles $\theta$ relative to the $c$-axis (inset schematic) exhibit prominent dHvA quantum oscillations. \textbf{e} FFT spectrum of $\Delta M_\tau$ (1/$H_\perp$) at $T = 0.4$ K and $\theta = 5.3^\circ$. Frequencies consistent with magnetic breakdown orbits are marked by stars. (inset) Expanded view of FFT spectrum at low $F \cos\left( \theta \right)$.

\pagebreak

\noindent\textbf{Commensurability oscillations in a stripe-modulated superlattice}
\\\textbf{a} $MR_\parallel\left( H \right)$ and \textbf{b} $MR_\perp\left( H \right)$ measured parallel and perpendicular to $\vec{q}$, respectively, in a FIB patterned ``L-bar" device at various fixed $T$ showing $1/H$ periodic oscillations. Oscillations persist in $MR_\parallel\left( H \right)$ to $T \approx 50$ K which are absent in $MR_\perp\left( H \right)$, consistent with commensurability oscillations (COs). \textbf{c} Schematic depiction of COs, which arise due to commensuration between the cyclotron diameter $2 r_\textnormal{c}$ and the modulation wavelength $\lambda$. \textbf{d} $\Delta MR_\parallel(1/H)$ at high $T$, extracted by subtracting a monotonic background from $MR_\parallel$ (dashed lines in \textbf{a}), showing commensurability oscillations with a characteristic $1/4$ period phase shift.

\pagebreak

\noindent\textbf{Superconductivity with suppressed interlayer coherence and in-plane anisotropy}
\\\textbf{a} Intralayer resistivity $\rho_\perp\left( T \right)$ measured perpendicular to the modulation $\vec{q}$ and interlayer resistivity $\rho_\textnormal{c}\left( T \right)$ show a separation in $T$ where they reach zero. \textbf{b} Temperature dependence of tunnel diode oscillator frequency shift $\Delta f(T)$ with $c$-axis (blue) and $ab$-plane (green) aligned with the coil axis $n$, respectively. Screening supercurrents, which are probed by $\Delta f$, onset at distinct $T$. 
\textbf{c} Zero-field cooled magnetic susceptibility $4\pi\chi \left( T \right)$ measured for $H = 5$ Oe aligned with the $c$-axis exhibits weak screening for $T_c < T < T^\star$, followed by a bulk Meissner state for $T < T_c$. The gray box delineates the mean and standard deviation of the normal state susceptibility, $\bar{\chi} (3 \textnormal{ K} < T < 5 \textnormal{ K}) \pm \sigma_\chi (3 \textnormal{ K} < T < 5 \textnormal{ K})$. \textbf{d} Depiction of separation of corresponding energy scales between intralayer and interlayer coherence due to mismatched 1D PDW SC. \textbf{e} Experimental configuration and \textbf{f} polar plot of interlayer critical current density $J_c$ (sample A) versus in-plane magnetic field orientation $H_{ab}(\varphi)$ exhibiting prominent two-fold anisotropy.

\end{document}